\documentstyle[12pt,epsfig]{article}
\def\jrn#1#2#3#4{{#1} {\bf #2} (#4) #3}
\def\PRL{Phys. Rev. Lett.}

\def\PRD{Phys. Rev. D}

\def\YaF{Yad. Fiz.}

\begin{document}
\pagestyle{empty}

\rightline{  }

\vspace*{1.5cm}

\begin{center}

\LARGE{Sleptons at post-WMAP benchmark points at LHC(CMS)\\[20mm]}

\large{Yu.M.Andreev \footnote{e-mail: andreyev@inr.ru},
S.I.Bityukov \footnote{e-mail: bityukov@mx.ihep.su} and
N.V.Krasnikov \footnote{e-mail: krasniko@ms2.inr.ac.ru}\\[8mm]}

\it{Institute for Nuclear Research RAS,\\
Moscow, 117312, Russia\\[15mm]}

\large{\rm{Abstract}} \\[7mm]

\end{center}

\begin{center}

\begin{minipage}[h]{14cm}
We study a possibility to detect sleptons at
post-WMAP benchmark points at LHC(CMS). We
find that at $L_{tot}=30 fb^{-1}$ it would be possible
to detect sleptons at points A, B, C, D, G.
We also investigate the production and decays of
right and left  sleptons separately.
We find that at $L_{tot}=30 fb^{-1}$ it would be possible
to detect right  sleptons with a mass up to
200 GeV and left  ones with a mass up to 300 GeV.
\end{minipage}

\end{center}

\newpage

\pagestyle{plain}

\section{Introduction}

One of the supergoals of the Large Hadron Collider (LHC) 
\cite{1} is the discovery of the supersymmetry. 
In particular, it is very important to investigate the possibility  of 
discovering nonstrongly   interacting superparticles (sleptons, higgsino, 
gaugino). In Refs. \cite{AgAm}-\cite{Baernew} sleptons discovery 
potential was
investigated for direct sleptons production via Drell-Yan mechanism and 
``generic'' LHC detector. In Ref.~\cite{Baernew} the production of sleptons 
from chargino and neutralino decays had been considered.
In Ref. \cite{Deneg} the LHC slepton discovery 
potential 
was investigated within the minimal supersymmetric model (MSSM) in the minimal 
supergravity (mSUGRA) scenario ($\tan \beta = 2$ case)
for Compact Muon Solenoid (CMS) detector. 
In Refs. \cite{Kr} - \cite{BiKr} the 
(LHC)CMS sleptons discovery potential and the possibility to discover 
lepton number violation in sleptons decays were investigated for direct 
production of right and left sleptons  within MSSM model.   

In this paper we investigate the possibility to discover sleptons 
at LHC(CMS) 
for post-WMAP supersymmetric benchmark scenarios \cite{WMAP}. These 
benchmark points  take into account WMAP   and 
other cosmological data, as well as the LEP and $b \rightarrow s \gamma$ 
constraints. We also reanalyze the LHC(CMS) 
discovery potential for the case of
direct production of right 
and left  sleptons in the MSSM model with arbitrary relation 
between the mass of lightest stable superparticle (LSP) and the 
slepton mass. 
One of the important ``technical'' differences between this paper and the 
previous studies is that we use PYTHIA program \cite{Pyth}  
for both simulation of 
background and signal supersymmetric events whereas in 
Refs.\cite{Deneg} -\cite{BiKr}
the PYTHIA program was used for 
the simulation of background events and ISAJET program \cite{Isajet} 
for simulation 
of supersymmetric events. As in Refs. \cite{Deneg}-\cite{BiKr} 
we use the CMS fast 
detector simulation program CMSJET \cite{Isajet}.    
We find that at total luminosity $L_{tot} = 30 fb^{-1}$ it would be possible
to detect sleptons at post-WMAP points   B, C, D, G.
We also find that at $L_{tot}=30 fb^{-1}$ it would be possible
 to detect right sleptons with a mass up to
 200 GeV and left ones with a mass up to 300 GeV. The organization of 
the paper is the following. In  Section 2 we review the main 
features of the mSUGRA model \cite{Msugra}  
and describe proposed in Ref.\cite{WMAP} 
post-WMAP benchmark points  for supersymmetry. In Section 3 we 
describe sleptons production 
mechanisms and sleptons decays relevant for this study. Section 4 is 
devoted to the discussion of the background and cuts used to 
suppress the background. In Section 5 we present the results 
of our numerical calculations. Section 6 contains concluding remarks. 
  
\section{ Post-WMAP benchmarks}

In the MSSM supersymmetry is broken at some high scale $M$ by generic 
soft terms so in general all soft SUSY breaking terms are arbitrary 
that complicates the analysis and spoils the predictive power of the theory. 
In mSUGRA model \cite{Msugra} the universality of different soft parameters 
at Grand Unified Theory (GUT) scale 
$M_{GUT} \approx 2 \cdot 10^{16} ~GeV$ is postulated. 
Namely, all the spin zero 
particle masses (squarks, sleptons, higgses) are postulated to be equal to 
the universal value $m_0$ at GUT scale. All gaugino particle masses 
are postulated to be equal to the universal value $m_{1/2}$ at GUT scale. 
Also the coefficients in front of quadratic and cubic SUSY soft breaking  
terms are postulated to be equal. The renormalization group equations are 
used to relate GUT and electroweak scales. The equations for the 
determination of nontrivial minimum of the electroweak potential are 
used to decrease the number of unknown parameters by two. So 
mSUGRA model depends on five unknown parameters. At present more or 
less standard choice of free parameters in mSUGRA model includes 
$m_0, m_{1/2}, \tan \beta , A$ and $ sign(\mu) $ \cite{Msugra}. 
All sparticle masses depend on these parameters.        
For instance, the slepton masses
of the first two generations are determined by the   
formulae \cite{Msugra}
\begin{equation}
m^2_{\tilde{l}_R} = m^2_0  + 0.15m^2_{1/2} -\sin^2\theta_{W}M^2_Z
\cos2\beta \,,
\end{equation}
\begin{equation}
m^2_{\tilde{l}_L} = m^2_0  + 0.52m^2_{1/2} -
1/2(1 -2\sin^2\theta_{W})M^2_Z \cos2\beta \,,
\end{equation}
\begin{equation}
m^2_{\tilde{\nu}} = m^2_0  + 0.52m^2_{1/2} + 1/2 \cos^2\theta_{W}M^2_Z
\cos2\beta \,.
\end{equation}
Charged left sleptons are the heaviest sleptons whereas the 
right sleptons are the lightest sleptons.  
For gaugino masses the following approximate formulae take place:
\begin{equation}
M_{\tilde{\chi}^0_1} \approx 0.45m_{1/2} \,,
\end{equation}  
\begin{equation}
M_{\tilde{\chi}^0_2} \approx M_{\tilde{\chi}^{\pm}_1} 
\approx   2M_{\tilde{\chi}^0_1}\,,
\end{equation}  
\begin{equation}
M_{\tilde{\chi}^0_2} \approx (0.25 - 0.35)M_{\tilde{g}} \,.
\end{equation}  
In mSUGRA model the $\tilde{\chi}^0_1$ gaugino is the lightest 
stable superparticle (LSP).

As it has been mentioned before in  mSUGRA model sparticle masses depend 
on five unknown 
parameters 
that complicates numerical analysis of the LHC SUSY discovery potential. 
In Ref.\cite{Bat} benchmark sets of supersymmetric parameters 
(13 post-LEP points)
 within mSUGRA model were suggested for further careful analysis.
The suggested points take into account the constraints from 
LEP, Tevatron, $b \rightarrow s \gamma$, $g_{\mu} -2 $ and cosmology.
Recently in Ref.\cite{WMAP} upgraded benchmark sets 
(post-WMAP benchmarks) were proposed. 
These  post-WMAP benchmarks 
take into account new WMAP data on dark matter density 
of the Universe. The mSUGRA model parameters 
and some sparticle masses
for these post-WMAP 
benchmark points are given in Table 1.

\begin{table}[htb] 
\begin{center}
\begin{tabular}{rrrrrrrrrrrrrr}
\hline
\hline
Point & $m_{1/2}$ & $m_0$ & tan$\beta$ & $sgn(\mu)$ & $A_0$ 
& $\tilde{\chi^0_1}$ &  $\tilde{\chi^0_2}$ & 
$\tilde{e}_L, \tilde{\mu}_L$ &    $\tilde{e}_R, \tilde{\mu}_R$ & 
 $\tilde{\nu}_e, \tilde{\nu}_{\mu}$ & $\tilde{\tau}_1$ & $ \tilde{\tau}_2$ &
$\tilde{\nu}_{\tau}$ \\
\hline
 A &  600 & 107 &  5 & + & 0 & 242 & 471 & 425 & 251 & 412 & 249 & 425 & 411 \\
 B &  250 &   57 & 10 & + & 0 & 95 & 180 & 188 & 117 & 167 & 109 & 191 & 167 \\
 C &  400 &   80 & 10 & + & 0 & 158 & 305 & 290 & 174 & 274 & 167 & 291 & 273 
\\
 D &  525 &  101 & 10 & $-$ & 0 & 212 & 415 & 376 & 224 & 362 & 217 & 376 & 
  360  \\
 E &  300 & 1532 & 10 & + & 0 & 112 & 184 & 1543 & 1534 & 1539 & 1521 & 1534 & 
 1532 \\
 F & 1000 & 3440 & 10 & + & 0 & 421 & 610 & 3499 & 3454 & 3492 & 3427 & 3485 & 
 3478 \\
 G &  375 &  113 & 20 & + & 0 & 148 & 286 & 285 & 185 & 270 & 157 & 290 & 266\\
 H &  935 &  244 & 20 & + & 0 & 388 & 750 & 679 & 426 & 665 & 391 & 674 & 657 
\\
 I &  350 &  181 & 35 & + & 0 & 138 & 266 & 304 & 227 & 290 & 150 & 312 & 278 
\\
 J &  750 &  299 & 35 & + & 0 & 309 & 598 & 591 & 410 & 579 & 312 & 579 & 558 
\\
 K & 1300 & 1001 & 39.6 & $-$ & 0 & 554 & 1064 & 1324 & 1109 & 1315 & 896 & 
1251 & 1239 \\
 L &  450 &  303 &  45 & + & 0 & 181 & 351 & 434 & 348 & 423 & 194 & 420 & 387 
\\
 M & 1840 & 1125 & 45.6 & + & 0 & 794 & 1513 & 1660 & 1312 & 1648 & 796 & 
1504 & 1492\\
\hline
\hline
\end{tabular}
\caption{ The mSUGRA parameters and some sparticle masses for 
proposed post-WMAP benchmarks (all masses in GeV), as calculated in 
ISASUGRA 7.67 (see Table 2 in Ref.\cite{WMAP}).}
\label{sets}
\end{center}
\end{table}


\section{Sleptons production and decays}

When sleptons are heavy relative to $\tilde{\chi}^{\pm}_1, 
\tilde{\chi}^{0}_1 $ sleptons are produced at the LHC only through 
Drell-Yan mechanism (direct slepton production), via $q\bar{q}$ annihilation 
with neutral or charged boson exchange in the s-channel, namely, 
$pp \rightarrow \tilde{l}_L \tilde{l}_L, \tilde{l}_R \tilde{l}_R, 
\tilde{\nu} \tilde{\nu}, \tilde{\nu} \tilde{l}, \tilde{l}_L\tilde{l}_R, 
$. The left sleptons decay to charginos and neutralinos via the 
following (kinematically accessible) decays:
\begin{equation}
\tilde{l}^{\pm}_L \rightarrow l^{\pm} + \tilde{\chi}^0_{1,2} \,,
\end{equation}
\begin{equation}
\tilde{l}^{\pm}_L \rightarrow \nu_{l}  + \tilde{\chi}^{\pm} \,,
\end{equation}
\begin{equation}
\tilde{\nu} \rightarrow \nu_{l}  + \tilde{\chi}^{0}_{1,2} \,,
\end{equation}
\begin{equation}
\tilde{\nu} \rightarrow l^{\pm}    + \tilde{\chi}^{\mp}_1 \,.
\end{equation}
For right sleptons only decays to neutralino are possible and they decay 
mainly to LSP:
\begin{equation}
\tilde{l}^{\pm}_R \rightarrow l^{\pm} + \tilde{\chi}^0_{1} \,,
\end{equation}
Note that an account of the 
mixing between  left and right charged sleptons slightly complicates 
the situation and allows decays (7,8) for eigenstates 
of $\tilde{l}_L$ and $ \tilde{l}_R$. 
If decays to second neutralino or first chargino are kinematically possible, 
the most interesting decays of $\tilde{\chi}^{\pm}_1, \tilde{\chi}^{0}_2$ 
are the following:
\begin{equation}
\tilde{\chi}^0_2 \rightarrow \tilde{\chi}^0_1 + l^{+}l^{-} \,,
\end{equation}
\begin{equation}
\tilde{\chi}^0_2 \rightarrow \tilde{\chi}^0_1 + \nu \bar{\nu} \,,
\end{equation}
\begin{equation}
\tilde{\chi}^0_2 \rightarrow \tilde{\chi}^0_1 + Z^{0} \,,
\end{equation}
\begin{equation}
\tilde{\chi}^{\pm}_1 \rightarrow \tilde{\chi}^0_1 + l^{\pm} + \nu \,,
\end{equation}
\begin{equation}
\tilde{\chi}^{\pm}_1 \rightarrow \tilde{\chi}^0_1 + W^{\pm} \,,
\end{equation}

If sleptons are light relative  to  $\tilde{\chi}^{\pm}_1, 
\tilde{\chi}^{0}_1 $ sleptons can be produced besides Drell-Yan mechanism 
from chargino and neutralino decays ($\tilde{\chi}^{\pm}_1, \tilde{\chi}^0_2$ 
indirect production), namely:
\begin{equation}
\tilde{\chi}^{0}_2 \rightarrow \tilde{l}^{\pm}_{L,R} l^{\mp} \,,
\end{equation} 
\begin{equation}
\tilde{\chi}^{0}_2 \rightarrow \tilde{\nu} \nu \,,
\end{equation} 
\begin{equation}
\tilde{\chi}^{\pm}_1 \rightarrow \tilde{\nu} l^{\pm}\,,
\end{equation} 
\begin{equation}
\tilde{\chi}^{\pm}_1 \rightarrow \tilde{l}^{\pm} \nu \,.
\end{equation} 

\section{Signature and background }

The slepton production and decays described in previous section lead to 
the signature  
with the simplest event topology: $ two ~leptons + E_{T}^{miss} + 
no ~jets $. This signature arises for both direct 
and indirect slepton pair production. 
In the case of indirectly produced sleptons not only event topology with two 
leptons but with single, three and four leptons are possible. Besides indirect 
slepton production from decays of squarks and gluino through charginos, 
neutralinos can lead to event topology   $ two ~leptons + E_{T}^{miss} + 
(n \geq 1) ~jets $. 

In this paper we use the event topology  
$ two~leptons + E^{miss}_T + no~jets$ 
to detect sleptons at LHC(CMS).  
Our simulations are made at the particle level with parametrized
detector responses based on a detailed detector simulation.
The CMS detector simulation program CMSJET 4.703 \cite{Abd} is used.
It incorporates the full ECAL and HCAL granularity. 
The energy resolutions for electrons (photons), 
 hadrons and jets are parametrized. Transverse 
and longitudinal profiles are also included according to parameterizations.

All the SUSY processes except particle spectrum are generated
with PYTHIA 6.152 \cite{Pyth}. Sparticle masses for updated post-WMAP 
benchmark points were taken from Ref.\cite{WMAP}. 
The Standard Model backgrounds are
also generated with PYTHIA 6.152. 
In our calculations we used the CTEQ 5L parton distribution set.
The signature used for the search
for sleptons at LHC is: two same-flavour opposite-sign leptons +
$E^{miss}_T + no ~jets$ \cite{AgAm}~-~\cite{BiKr}. 
Our cuts are the following:

$ $

a. for leptons:
 
\begin{itemize}
\item { 
  $p_T$ - cut on leptons ($p_T^{lept} \geq p_T^{lept,0}$) 
and lepton isolation within ${\Delta}R<0.3$ cone
  with ISOL$<$0.1 (CMSJET default); }
\item {
  effective mass of two opposite-sign leptons of the same flavour: outside
  $M_Z \pm {\delta}M_Z$ band (${\delta}M_Z$ = 10 GeV); }
\item { $\Delta \Phi(l^{+}l^{-})  < \Delta \Phi_{ll}^{0}$  cut; }

b. for $E^{miss}_T$ :

\item { $E_T^{miss}> E_T^{miss,0} $ cut; }
\item { $\Delta \Phi$ ( $E_T^{miss}$, $ll)$  $> 
\Delta \Phi^{0}$  cut  for relative azimuthal angle 
 between two same-flavour opposite sign leptons}

c. for jets:

\item { jet veto cut: $N_{jet} = 0$ for some $E^{jet}_T > E^{jet, 0}_T$ 
threshold in  pseudorapidity interval $|\eta_{jet}| < 4.5$}.
\end{itemize}

Such type of cuts is the standard one and it was used in previous 
Refs.\cite{AgAm}~-~\cite{BiKr}.

In this paper we use the set of 10 cuts, see Table 2.\\
\begin{table}[htb]
\begin{center}
\begin{tabular}{rrrrrrr}
\hline
\hline
      & $p_T^{lept,0}$ & $E_T^{miss,0}$ &  $\Delta \Phi_{ll}^0$
      &$E_T^{jet,0}$   & ${\delta}M_Z$ & $\Delta \Phi^{0}$\\
\hline
Cut1 & 20 & 50 &  130 &  30 & 10 & 160 \\
Cut2 & 20 & 50 &  $-$ &  30 & 10 & 160 \\
Cut3 & 50 &140 &  140 &  60 & 10 & 150 \\
Cut4 & 50 &100 &  130 &  30 & 10 & 150 \\
Cut5 &100 &200 &  130 &  60 & 10 & 150 \\
Cut6 & 60 &150 &  130 &  45 & 10 & 150 \\
Cut7 & 80 &120 &  140 &  70 & 10 & 145 \\
Cut8 & 75 &170 &  160 & 100 & 10 & 160 \\
Cut9 & 30 &75 &  130 & 45 & 10 & 150 \\
Cut10 &40 &90 &  130 & 50 & 10 & 150 \\
\hline
\hline
\end{tabular}
\caption{ The parameters of the used cuts.}
\label{sets}
\end{center}
\end{table}


The main Standard Model (SM) backgrounds are: $WW$, 
$WZ$, $Wt\bar{b}$, $t\bar{t}$,
$\tau\bar{\tau}$, $b\bar{b}$.
The distributions of the SM background on $p^{lept}_T$ and $E_T^{miss}$ 
are presented in Figs.1-4.
The contribution of $WW$ background is
(40-80)\% in the dependence  on the cut number.
There are also internal SUSY backgrounds which arise through 
$\tilde{q}\tilde{q}$, $\tilde{g}\tilde{g}$ and $\tilde{q}\tilde{g}$ 
productions and subsequent cascade decays with jets outside 
acceptance or below threshold. SUSY backgrounds depend on 
SUSY masses and as a rule they are small compared to SM backgrounds.
Note that when we are interested in new physics discovery (the first stage of 
any data analysis) we have to compare the calculated number of 
standard background 
events $N_{bg}$ with new physics signal events $N_{new~physics} = N_{slept} + 
N_{susy, bg}$,  so SUSY background events increase the discovery potential 
of new physics.   

\begin{figure}[htpb]
\begin{center}
\epsfig{file=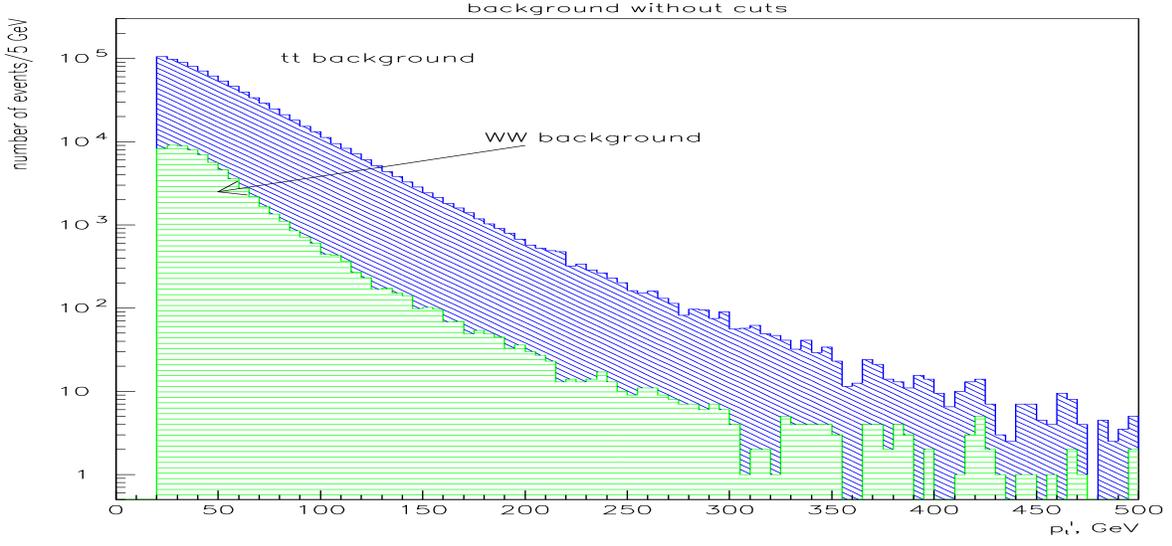, width=17.cm, height=8.cm}
\caption{Leptons $p^{lept}_T$ distributions for main 
SM background (WW, $t \bar t$) before any cuts ($L_{tot}=10~fb^{-1}$).}
    \label{fig:1} 
\end{center}
\end{figure}

\begin{figure}[htpb]
\begin{center}
\epsfig{file=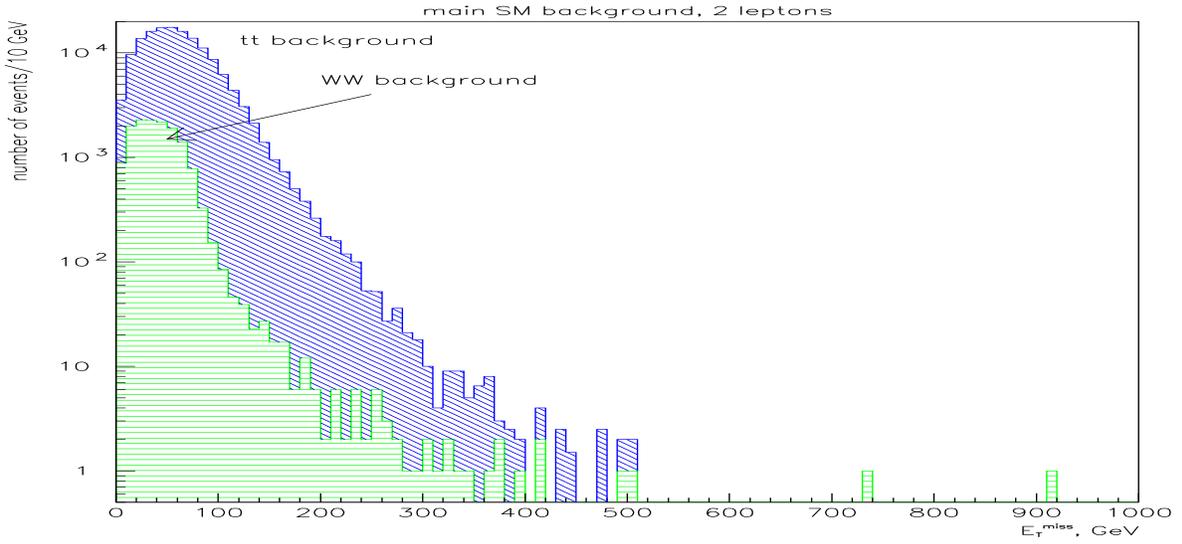, width=17.cm, height=8.cm}
\caption{$E_T^{miss}$ distributions for main SM background (WW, $t \bar t$)
events with two isolated leptons $p^{lept}_T>20~GeV$
($L_{tot}=10~fb^{-1}$).}
    \label{fig:2} 
\end{center}
\end{figure}

\begin{figure}[htpb]
\begin{center}
\epsfig{file=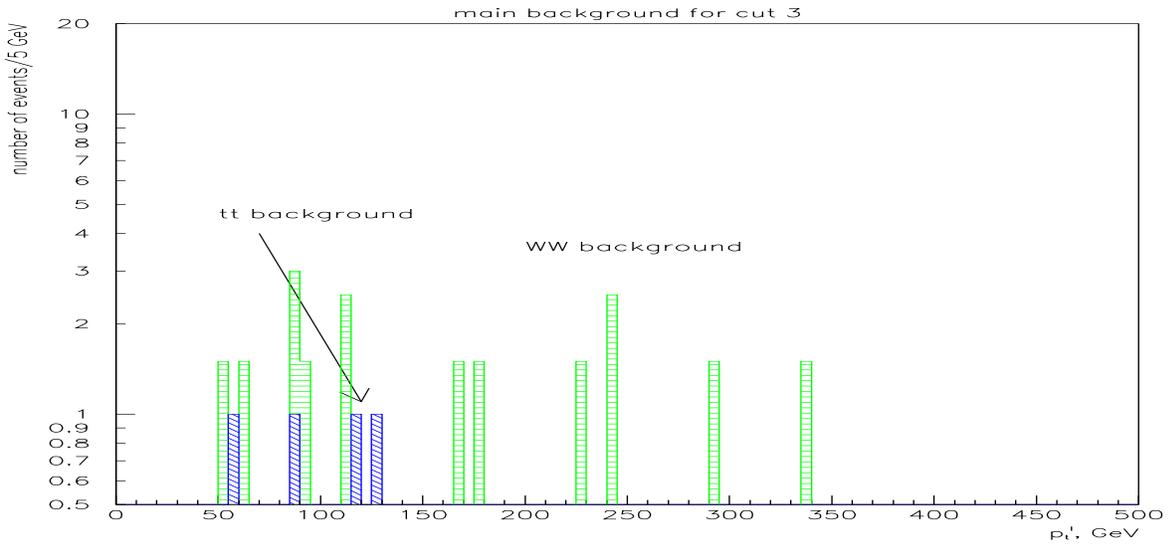, width=17.cm, height=8.cm}
\caption{Leptons $p^{lept}_T$ distributions for main SM background (WW, $t \bar t$)
for cut 3 ($L_{tot}=10~fb^{-1}$).}
    \label{fig:3} 
\end{center}
\end{figure}

\begin{figure}[htpb]
\begin{center}
\epsfig{file=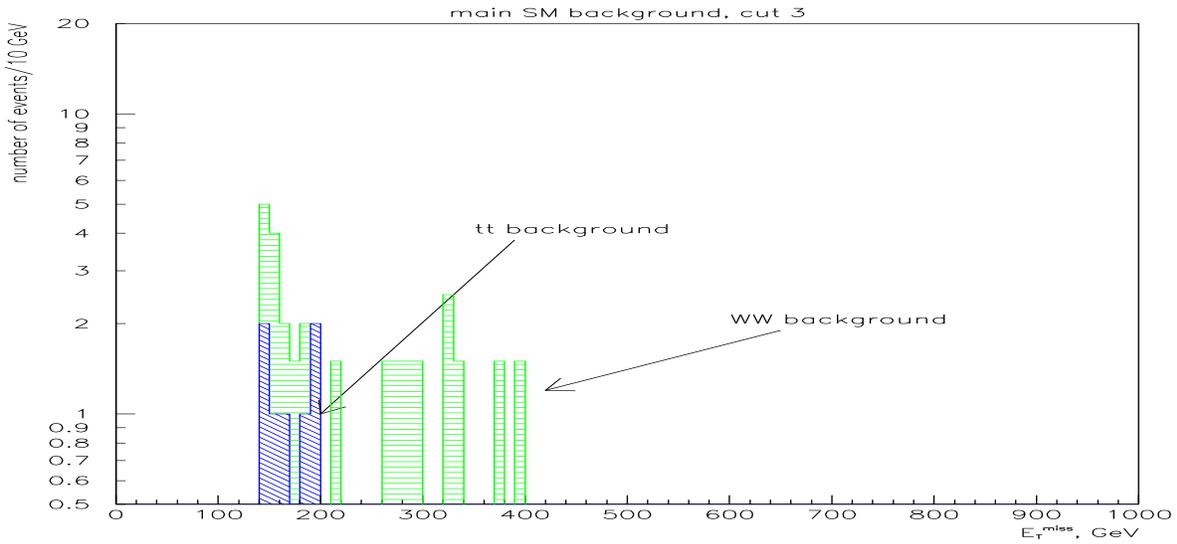, width=17.cm, height=8.cm}
\caption{$E_T^{miss}$ distributions for main SM background (WW, $t \bar t$)
events for cut 3 ($L_{tot}=10~fb^{-1}$).}
    \label{fig:4} 
\end{center}
\end{figure}

SM background cross sections after cuts are 
given in Table 3.(in fb)\\

\begin{table}[htb]
\begin{center}
\begin{tabular}{lcccccccccc}
\hline
\hline
Cut & 1 & 2 & 3 & 4 & 5 & 6 & 7 & 8 & 9 & 10 \\
\hline
$\sigma_{bg}$& 288 & 775 & 3.6 & 6.7 & 0.68 & 1.9 & 3.3 & 3.0 & 101 & 24 \\
\hline
\hline
\end{tabular}
\caption{ The SM background cross sections after cuts (in $fb$).}
\label{sets}
\end{center}
\end{table}

\section{Results}

For post-WMAP points (A - M) our results are the following.
We  found that at $L_{tot}=10 fb^{-1}$ it would be possible
to discover sleptons only at point B.\footnote{In our calculations we used the 
approximate formula for the
significance  $S = \frac{2N_S}{\sqrt{N_B} + 
\sqrt{N_S + N_B}}$ that is appropriate characteristic for 
future experiments, see Refs.\cite{KB} and also Ref.\cite {Bar}. }  
For cut 3 we found that $N_S = 45$, $N_B = 38$, $ S = 5.9$.
For  $L_{tot}=30 fb^{-1}$ it is possible to discover sleptons at 
points $B, C, D, F$, see Table 4.
\footnote{ See also Figs.5-6 for an illustration of the dependence of 
the  background and the signal on the  cut parameters} .  

\begin{figure}[htpb]
\begin{center}
\epsfig{file=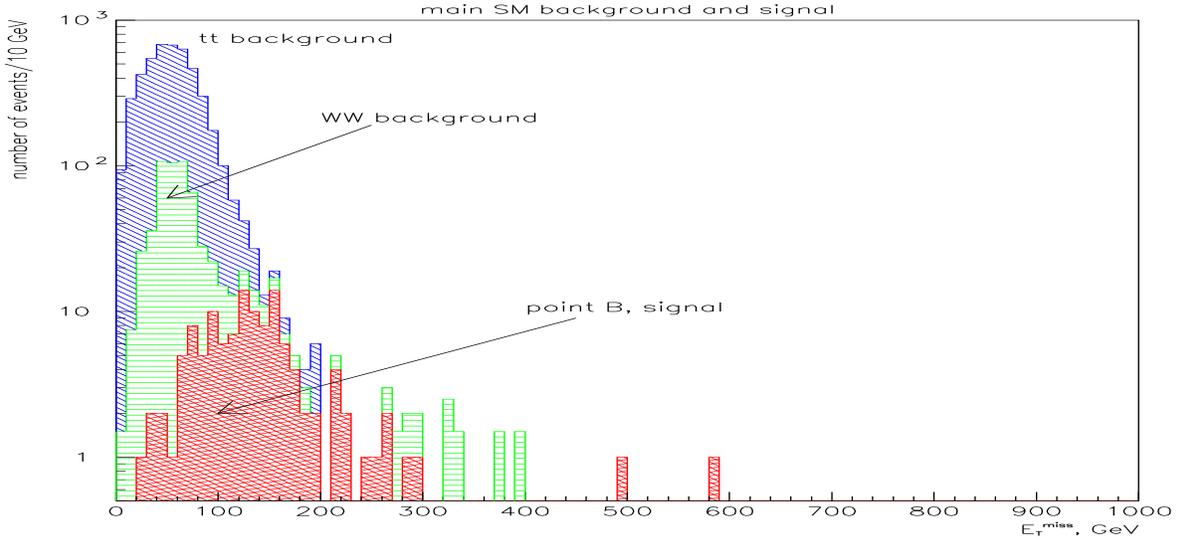, width=17.cm, height=8.cm}
\caption{$E_T^{miss}$ distributions for main SM background (WW, $t \bar t$)
and signal at point B for events with two isolated leptons $p_T^{lept}>50~GeV$
for cut 3 before cuts on $E_T^{miss}$ and $\Delta \Phi^0$
($L_{tot}=10~fb^{-1}$).}
    \label{fig:5} 
\end{center}
\end{figure}

\begin{figure}[htpb]
\begin{center}
\epsfig{file=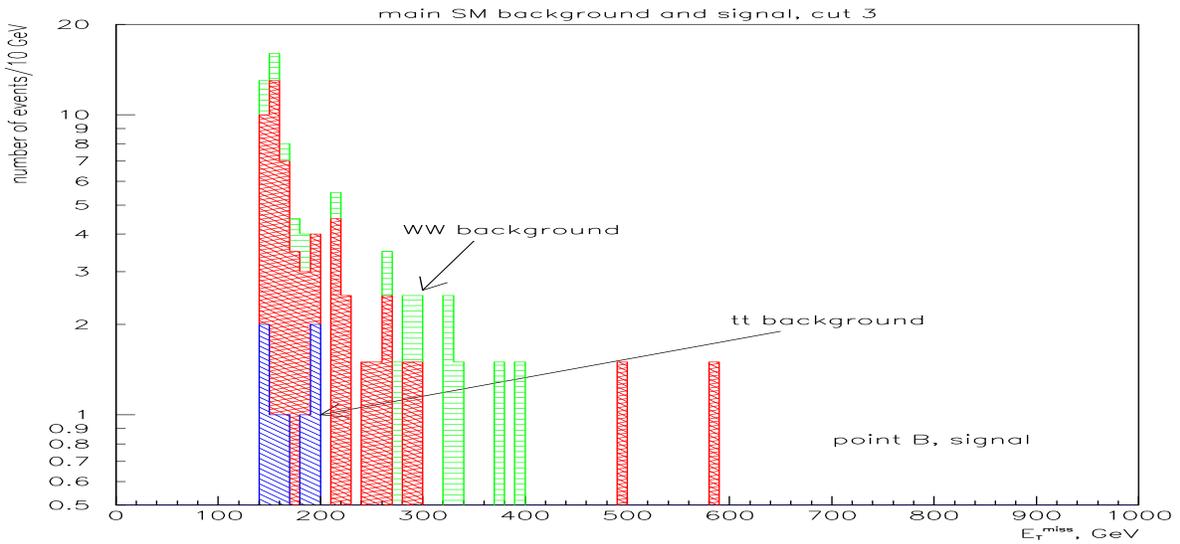, width=17.cm, height=8.cm}
\caption{$E_T^{miss}$ distributions for main SM background (WW, $t \bar t$)
and signal at point B for cut 3 ($L_{tot}=10~fb^{-1}$).}
    \label{fig:6} 
\end{center}
\end{figure}

\begin{table}[htb]
\begin{center}
\begin{tabular}{ccccc}
\hline
\hline
Point & Cut & $N_S$ & $N_B$ & S\\
\hline
B     &  4  &  180  &  212 & 10.5 \\
C     &  3  &   84  &  112  & 6.8  \\
D     &  3  &   61  &  110 & 5.2 \\
G     &  6  &   49  &  57 & 5.5 \\
\hline
\hline
\end{tabular}
\caption{ Sleptons discovery points at $L_{tot} = 30~fb^{-1}$}
\end{center}
\end{table}

At  $L_{tot} = 100 fb^{-1}$ the sleptons  
discovery points are A,B,C,D,G,I. \footnote{We did not take into account 
pileup effects therefore the results for 
high luminosity $L_{tot} = 100~fb^{-1}$ are rather preliminary. 
We think that  the  use of ``hard'' cuts 3 - 8 allows to minimize the 
influence of pileup effects on the significance.}  
We also investigated the slepton discovery potential for 
post-LEP benchmark points \cite{Bat} and found that the LHC(CMS) 
slepton discovery potential  for post-LEP points coincides with the 
slepton discovery potential for post-WMAP points.  

In this paper we  studied also the production and decays of
right and left sleptons separately.\footnote{To be precise we considered the 
production and decays of the 
first and second generation sleptons $\tilde{e}_R, \tilde{e}_L, 
\tilde{\nu}_{e_L}, \tilde{\mu}_R, \tilde{\mu}_L, 
\tilde{\nu}_{{\mu}_L}$. An account of the third generation sleptons 
with the masses equal to the masses of the first and second generation 
sleptons is not essential since $Br(\tau \rightarrow ~leptons) \approx 0.35$.}
In this study we  assumed
that sleptons decay mainly into LSP and leptons:
\begin{equation}
\tilde{l}_R^{-} \rightarrow l^{-} + \tilde{\chi}_{1}^0 \\,
\end{equation}
\begin{equation}
\tilde{l}_L^{\pm} \rightarrow l^{\pm} + \tilde{\chi}_{1}^0\\.
\end{equation}
Of course, in real life we expect that the decays of other sparticles 
will also contribute to the signature  
$ two ~leptons + E^{miss}_T + no ~jets$. But if we are interested 
in new physics signal discovery  additional 
contribution only increases new physics discovery potential of this signature.
 
We  made simulations for LSP mass $m_{LSP}$ equal
to $0.2~m_{\tilde l}$,
$0.4~m_{\tilde l}$, $0.6~m_{\tilde l}$ and $0.8~m_{\tilde l}$.\footnote{ We 
assume that $m_{\tilde{e}_R} = m_{\tilde{\mu}_R}$ and 
 $m_{\tilde{e}_L} = m_{\tilde{\mu}_L}$}.
The dependence of the cross section for the production of right and left 
sleptons for the case  of two flavour degenerate right and left charged 
sleptons is presented in Fig.7.  
Our results are given  in Table 5.\\

\begin{figure}[htpb]
\begin{center}
\epsfig{file=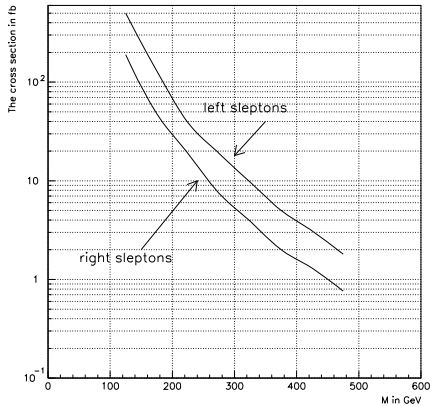,width=6.2cm}
\caption{Cross section $\sigma(pp \rightarrow \tilde l_R \tilde l_R$)
for various values of the right slepton masses and cross section 
$\sigma(pp \rightarrow \tilde l_L \tilde l_L$)
for various values of the left slepton masses at LHC (in fb).}
    \label{fig:7} 
\end{center}
\end{figure}

\begin{table}[htb]
\begin{center}
Left  sleptons\\
$L=10 fb^{-1}$
\begin{tabular}{c|cccccccc}
\hline
\hline
  0.8 & - & - & - & - & - & - & - & - \\
  0.6 & + & - & - & - & - & - & - & - \\
  0.4 & + & + & - & - & - & - & - & - \\
  0.2 & + & + & + & - & - & - & - & - \\
\hline
$m_{LSP}/m_{\tilde l_{R}}$&100&150&200&250&300&350&400&450\\
\hline
\end{tabular}

$L=30 fb^{-1}$
\begin{tabular}{c|cccccccc}
\hline
  0.8 & - & - & - & - & - & - & - & - \\
  0.6 & + & + & + &  & - & - & - & - \\
  0.4 & + & + & + & + & + & - & - & - \\
  0.2 & + & + & + & + & + & - & - & - \\
\hline
$m_{LSP}/m_{\tilde l_{R}}$&100&150&200&250&300&350&400&450\\
\hline
\end{tabular}

$L=100 fb^{-1}$
\begin{tabular}{c|cccccccc}
\hline
  0.8 & - & - & - & - & - & - & - & - \\
  0.6 & + & + & + & + & + & - & - & - \\
  0.4 & + & + & + & + & + & + & - & - \\
  0.2 & + & + & + & + & + & + & + & - \\
\hline
$m_{LSP}/m_{\tilde l_{R}}$&100&150&200&250&300&350&400&450\\
\hline
\hline
\end{tabular}

Right  sleptons\\
$L=10 fb^{-1}$
\begin{tabular}{c|ccccccc}
\hline
\hline
  0.8 & - & - & - & - & - & - & -\\
  0.6 & - & - & - & - & - & - & -\\
  0.4 & - & - & - & - & - & - & -\\
  0.2 & + & - & - & - & - & - & -\\
\hline
$m_{LSP}/m_{\tilde l_{L}}$&100&150&200&250&300&350&400\\
\hline
\end{tabular}

$L=30 fb^{-1}$
\begin{tabular}{c|ccccccc}
\hline
  0.8 & - & - & - & - & - & - & -\\
  0.6 & - & - & - & - & - & - & -\\
  0.4 & + & - & - & - & - & - & -\\
  0.2 & + & + & + & - & - & - & -\\
\hline
$m_{LSP}/m_{\tilde l_{L}}$&100&150&200&250&300&350&400\\
\hline
\end{tabular}

$L=100 fb^{-1}$
\begin{tabular}{c|ccccccc}
\hline
  0.8 & - & - & - & - & - & - & -\\
  0.6 & + & - & - & - & - & - & -\\
  0.4 & + & + & + & + & - & - & -\\
  0.2 & + & + & + & + & + & - & -\\
\hline
$m_{LSP}/m_{\tilde l_{L}}$&100&150&200&250&300&350&400\\
\hline
\hline
\end{tabular}
\caption{The right and left sleptons LHC(CMS) 
$5~\sigma$ discovery potential for different luminosities.}
\end{center}
\end{table}


As it follows from our results the sleptons discovery potential depends
on the LSP mass. For $m_{LSP} = 0.2~m_{\tilde{l}}$ it would be possible
to detect right sleptons with a mass up to
200 GeV and left ones with a mass up to 300 GeV. For instance, for 
right slepton with a mass $m_{\tilde{l}_R} =  200~GeV$ and LSP with
a mass $m_{LSP} = 40~GeV$ we found that $N_S = 70$, $N_B = 108$ , $S = 5.9$ 
( cut 3, $L_{tot} = 30~fb^{-1}$).  For  
left slepton with a mass $m_{\tilde{l}_R} =  200~GeV$ and LSP with
a mass $m_{LSP} = 40~GeV$ we  found that $N_S = 140$, $N_B = 108$ , $S = 10.7$ 
( cut 3, $L_{tot} = 30~fb^{-1}$).

\section{Conclusion}
 
In this paper we studied the possibility to detect sleptons at LHC(CMS).
For post-WMAP benchmark points we found that it is possible to 
discover sleptons at point $B$, points $B, C, D, F$ and points 
$A, B, C, D, G, I$ for total luminosities $L_{tot} = 10~fb^{-1}$,
 $L_{tot} = 30~fb^{-1}$ and  $L_{tot} = 100~fb^{-1}$ correspondingly.
 We also investigated the possibility to detect sleptons for the case 
when they decay dominantly to leptons and LSP. \footnote{For right sleptons 
they really decay mainly to leptons and LSP while for left sleptons 
for $m_{\tilde{l}} > m_{\tilde{\chi}^0_2}$ cascade decays can dominate.}
For $L_{tot} = 30~fb^{-1}$ we found that it is possible to discover right 
sleptons with masses up to $200~GeV$ and left sleptons with masses up to 
$300~GeV$.  

We are indebted to V.A.Matveev for valuable comments. The work of S.I.B.
and N.V.K. has been supported by RFFI grant No 03-02-16933.

\end{document}